\documentstyle[preprint,aps,epsfig]{revtex}

\begin{document}

\title{\bf Anisotropic spin and charge transport in 
Rashba Hamiltonian.}

\author{T. P. Pareek 
\\
Max-Planck-Institute f\"ur Mikrostrukturphysik, Weinberg 2,
D-06120 Halle, Germany}
\maketitle
\begin{abstract}
We explore spin and charge transport phenomena in two dimensional
electron gas in presence of Rashba spin-orbit coupling connected
to two ideal Ferromagnetic leads. In particular we show through a
combination of analytical and numerical calculation that the
spin polarization which is transported depends on the Magnetization
direction of ferromagnet even if the magnetization of both
FM's are parallel.Conductance is also shown to be anisotropic.
These anisotrpies present in spin and charge transport are
a consequence of breaking of rotational invariance due to Rashba spin-orbit
interaction and are present irrespective of the Hamiltonian considered being
an effective mass Hamiltonian or tight binding model Hamiltonian. 

\end{abstract}

PACS numbers: 72.25-b,72-25.Dc,72.25Mk,72.25Rb,72.25Hg

The growing field of spintronics has attracted a lot of interest after the
proposal of spin-filed effect transistor (SPIN-FET) by 
Datta and Das\cite{Datta}. The Datta-Das SPIN-FET is a hybrid structure
of type FM1-2DEG-FM2, where 2DEG is a  
two-dimensional electron gas of a narrow gap semiconductor (InAS)
and FM1 and FM2 are injector and detector Ferromagnetic contacts.
The working of this device relies on the manipulation of electronic spin state
in 2DEG with the electric field of an external gate electrode. Essential for
this mechanism is field dependent spin-orbit coupling, which is
relatively large and well established\cite{lommer}. It is now generally accepted that
the spin-orbit coupling in narrow-gap 2DEG is governed by Rashba Hamiltonian
\cite{rashba}. For a 2DEG lying in {\it xy} plane(see Fig.1) the
Rashba spin-orbit interaction has the form,
$H_{R}=\alpha ({\bf k}\times {\bf \sigma})\cdot \hat{z}$,with
${\bf k}$ being momentum vector, ${\bf {\sigma}}$ Pauli matrices and
$\hat{z}$ is a unit vector perpendicular to 2DEG plane. 
The Rashba spin-orbit causes spin splitting for ${\bf k} \neq 0$, 
$\Delta E = 2\alpha k$, which is linear in momentum.
The Rashba splitting is due to lack of space inversion
symmetry and not due to lack of time reversal symmetry. Since
Rashba Hamiltonian is time reversal invariant. However the
exchange splitting in Ferromagnets is due to the breaking
of time reversal symmetry. Therfore it is natural to expect that
spin and charge transport properties of a hybrid structure like
SPIN-FET, which combines elements with different symmetry properties,
may be different than the standard mesoscopic structures consisting of
elements with same symmetry, for, e.g., all metal mesoscopic structures.

Motivated by this, in this paper we study the spin and charge transport of a
FM1-2DEG-FM2 system shown in FIG. 1. The question we
are addressing is the following: Consider the Fig.1, a natural
reference frame for the Fig.1 is defined by the plane of 2DEG
(we call it {\it xy} plane) and the normal to this plane ,{ i.e.}, the
{\it z } axis. The polarization of the Ferromagnets FM1 and FM2 are parallel to
each other and 
points in a direction $(\theta,\phi)$ with respect to the natural 
coordinate system {\it i.e.} make an angle $\theta$ with
{\it z} axis and an angle $\phi$ with the {\it x } axis.Now the question 
is does the spin polarization which is transported from FM1 to FM2
and the charge transport, i.e, conductance
depends on $(\theta, \phi)$? 
In other words if we rotate 
the polarization vector of Ferromagnets simultaneously
with respect to the natural coordinate system in such a way that they
always remains parallel, does conductance and spin polarization
which is  transported changes?
Naively speaking one would expect that the
conductance and transported spin-polarization 
should be independent of $(\theta, \phi)$ as long as both
the Ferromagnets are parallel. In contrast to naive expectation we
show through a combination of  analytical and
numerical calculation  that spin polarization which is transported
and charge
conductance are anisotropic and these
anisotropies are present irrespective of
the Hamiltonian considered being an effective mass Hamiltonian\cite{Mole} or
tight binding Hamiltonian\cite{tribhu1}. 
In this sense this is a rather general principal which says the
polarization of transported electron across a FM/2DEG/FM2 and conductance
is anisotropic and is a 
consequence of breaking of rotational invariance due to spin-orbit coupling.
This is in contrast to the claim 
made by Molenkamp et. al.\cite{Mole} that the effective mass Hamiltonian
does not have conductance anisotropy while the tight binding model
\cite{tribhu1} has
due to reduced symmetry of lattice. Another important consequence of
our study is it points out that spin coherence is also anisotropic, {\it i.e.}
it depends on the chosen basis. 

The Hamiltonian of a 2DEG lying in {\it xy} plane (as shown
in Fig. 1), in presence of 
Rashba spin-orbit coupling reads:
\cite{rashba}
\begin{equation}
H=-\frac{\hbar {\bf \nabla}^2}{2m}+\alpha ({\bf \sigma} \times {\bf k})
\cdot {\bf \hat{z}}
\label{eq1}
\end{equation}
where $\alpha$ Rashba spin-orbit interaction parameter and
${\bf \sigma}$=($\sigma_{x},\sigma_{y},\sigma_{z}$) is the
vector of Pauli matrices and $\hat{z}$ is unit vector along the {\it z} axis.
We write the above Hamiltonian in the matrix form which is more convenient for
the study of spin transport
\begin{equation}
H=\frac{1}{2}(B_{0}{\bf I}+{\bf B_{R}} \cdot {\bf \sigma})
\label{eq2}
\end{equation}
where ${\bf I}$ is the $2\times 2$ identity matrix,
 $B_{0}$=$\frac{\hbar^{2}(k_{x}^2+k_y^{2})}{2m}$ and
the vector is ${\bf B_{R}}$=$ 2 \alpha (k_{y}
{\bf \hat{x}}-k_{x}{\bf \hat{y}})$.

An appropriate physical quantity to study the spin transport
is the Polarization vector ${\bf P}$=$<{\bf \sigma}>$ where
angular brackets represents the ensemble averaging. With
this definition one can immediately write down the equation of
motion for polarization vector,
\begin{eqnarray}
\frac{d{\bf P}}{dt}&=&\frac{d{\bf <\sigma>}}{dt}=-\frac{i}{\hbar}<{\bf \sigma}H
-H{\bf \sigma}> \nonumber \\
&=&\frac{1}{2i\hbar} <{\bf \sigma}(\bf B_{R} \cdot \sigma)
-(\bf B_{R} \cdot \sigma){\bf \sigma}>
\label{eq3}
\end{eqnarray}
simplifying above equation using vector identities for triple product
and commutation relation for Pauli matrices leads to following equation of motion for polarization vector,
\begin{equation}
\frac{d{\bf P}}{dt}={\bf B_{R}}\times <\bf \sigma> .
\label{eq4}
\end{equation}
The eq. (\ref{eq4}) is well know in the literature and is 
a fully quantum mechanical and holds eve if ${\bf B_{R}}$ is time dependent.
The eq. (\ref{eq4}) can be solved analytically when the field ${\bf B_{R}}$
is a constant vector, the most general solution is given as;
\begin{eqnarray}
{\bf P}(t)={\bf P_{0}}\cos(\omega_{R}t)+2 \hat{{\bf B_{R}}} (\hat{{\bf B_{R}}}
\cdot \hat{\bf P_{0}}) \sin^{2}(\omega_{R}t/2) \nonumber\\
+(\hat{{\bf B_{R}}}
\times \hat{\bf P_{0}})\sin(\omega_{R}t)
\label{eq5}
\end{eqnarray}
where $\bf P_{0}$ is the initial polarization imposed by
Ferromagnet FM1 (we are interested in the case when the Polarization of
both the Ferromagnets FM1 and FM2 are parallel and equal in magnitude,
{\it i.e}, $\bf P_{1}=P_{2}=P_{0}$) , $\omega_{R}=B_{R}/\hbar $ is precession frequency
(precession angle $\phi=\omega_{R}t$), 
$B_{R}=2\alpha\sqrt{k_{x}^2+k_{y}^2}$ is the magnitude of 
Rashba field ${\bf B_{R}}$( the direction of
${\bf B_{R}}$ is always perpendicular to the instantaneous
wave vector ${\bf k}$). During electrons free flight the direction and magnitude
of ${\bf B_{R}}$ remains constant hence the solution provided by 
eq.(\ref{eq5}) is applicable only during the free flight.
Since scattering from impurity or boundary changes the 
momentum and hence the filed $\bf B_{R}$, so the time
occurring in eq.(\ref{eq5}) is free flight time.
However for the ballistic transport we will use the eq.(\ref{eq5}) and take into account the boundary scattering later in diffusive approximation
as we will see later. 
Now since we are interested in the transport properties when the 
polarization of both the Ferromagnets are parallel to each other
but pointing in arbitrary direction ($\theta, \phi$) such that
${\bf P_{1}=P_{2}}={\bf P_{0}}=P_{0}(\sin\theta\cos\phi,
\sin\theta\sin\phi,\cos\theta)$ 
such that  with respect to
hence by projecting 
${\bf P}(t)$ on ${\bf P}_{0}$ we get the,
\begin{equation}
{\bf P}(t)\cdot{\bf P}_{0}=|{\bf P}_{0}|^2\cos(\omega_{R}t)+
({\bf P}_{0}\cdot\hat{\bf B_{R}})^2\sin^2(\omega_{R}t/2) ,
\label{eq6}
\end{equation}
\noindent where $\omega_{R} = B_{R}/\hbar \equiv 2\alpha k_{f}/\hbar$.
The eq.(\ref{eq6}) is the quantitative measure of spin polarization which 
gets transported through the 2DEG from FM1 to FM2.
For a given injection angle $\beta$ as shown in Fig. 1, the
eq.(\ref{eq6}) simplifies to,
\begin{eqnarray}
Pol(\theta,\phi,L,W,\omega_{R}t) &\equiv& \frac{{\bf P}(t)\cdot{\bf P}_{0}}{|{\bf P}_{0}|^2} \nonumber\\
=\cos(\omega_{R}t)+
\sin(\beta - \phi)^2\sin(\theta)^2\sin^2(\omega_{R}t/2) 
\label{eq7}
\end{eqnarray}
In the above equation $t$ is the time electron takes to reach the
output terminal and it is clear from eq.(\ref{eq7}) that value
of transported polarization ,i.e, $Pol(\theta,\phi,L,W,\omega_{R}t)$, 
lies between +1 and -1. 
Since the electron are injected over $-\pi/2 \le \beta \le \pi/2$, we need to make an average over all possible values of injection angle
$\beta$. However depending upon injection angle $\beta$ electron reaches the boundary without scattering(dashed trajectory in Fig. 1) or with scattering
(solid trajectory in Fig.1) from the boundaries.
Hence we need to calculate $t$ accordingly for different values of $\beta$.
Therefore we divide the the integration over $\beta$ in three regimes,
namely, (a) $-\pi/2 \le \beta \le -\tan^{-1}(W/2L) $,
(b) $-\tan^{-1}(W/2L)\le \beta \le -\tan^{-1}(W/2L) $ and (c) 
$\tan^{-1}(W/2L)\le \beta \le \pi/2$ where (a) and (c) corresponds to
the trajectories which suffers scattering from boundary while trajectories
in regime (b) propagates ballistically.
For injection angle $\beta$  in the regime (b)
electron reaches the
output terminal(FM2) ballistically therfore the time to
reach the output terminal is $t=L/\cos(\beta)$(see Fig.1 dashed line),
while for regime (b) and (c) 
electron scatters from the
boundary at least once before reaching the out put terminal(FM2), 
hence for these
values of $\beta$ we assume that the electrons diffuse along the
channel with a mean free path $W/2\sin(\beta)$ (later in our exact
numerical simulation we will see that this approximation is
quite reasonable) so the
time to reach the boundary is given as $t=(2 {L^2}\sin(\beta))/(v_{f}W)$.
Using these vale for $t$ we get
\begin{eqnarray}
\omega_{R}\,t=\left\{
\begin{array}{ll}
{\frac{2\alpha k_{f}L}{v_{f}\cos(\beta)} \equiv 
\frac{2\pi \tilde{\alpha} \tilde{L}}{\cos(\beta)}}&
\hspace{-1.5cm}\mbox{\hspace{0.1cm} $\beta$ $\in$ \{$-\tan^{-1}(\frac{W}{2L})$, 
$\tan^{-1}(\frac{W}{2L})$\}}  \\ 
{\frac{2\alpha k_{f}L^{2}\sin(\beta)}{v_{f}W} \equiv 
{\frac{4\pi\tilde{\alpha}\tilde{L^{2}}\sin(\beta)}{\tilde{W}}}}
& \mbox{\hspace{-0.2cm} {$\beta$ $\in$ 
\{$\pm{\frac{\pi}{2}}$, $\pm\tan^{-1}(\frac{W}{2L})$\}}}
\end{array}
\right.
\label{eq8} 
\end{eqnarray}
where ${\tilde{\alpha}=\alpha k_{f}/E_{f}}$ is dimensionless Rashba
parameter($E_{f}$ is Fermi energy), 
$\tilde{L}=L/\lambda_{f}$ and $\tilde{W}=W/\lambda_{f}$ is 
length and width of channel in units of Fermi wavelength.
Substituting these values of $\omega_{R}t$ in eq.(\ref{eq7}) 
and performing the integration over $\beta$ we obtain 
polarization as function of $\theta,\phi,\tilde{L},\tilde{W},\tilde{\alpha}$.
Eq.(\ref{eq7}) together with eq.(\ref{eq8}) can be used to calculate the
transported polarization for any given direction ($\theta,\phi$), however
for clarity and simplicity we
present results for three specific cases corresponding to
different values of $\theta$ and $\phi$, namely,
(i)$\theta$=0, $\phi$ is variable, i.e., polarization of FM1 nad FM2 is rotated
in {\it xy} plane (the plane formed by 2DEG) 
(ii) $\phi$=0, $\theta$ is variable
corresponding to the rotation in {\it xz} plane 
(iii)$\phi$=$\pi/2$, $\theta$ is
variable, corresponding to the rotation in {\it yz} plane.
For these three different cases the transported polarization
given by eq.(\ref{eq7})
is shown in Fig.2 as a function of angle , where the other parameters are 
$\tilde{L}=\tilde{W}=50/(2\pi)$ and $\tilde{\alpha}=0.06$. It is
clearly seen from Fig.2 that polarization which is transported is anisotropic,
it is a consequence of spin-orbit 
coupling which breaks the rotational symmetry.
The amplitude of oscillation tells about the spin coherence and since
this is different for all the three cases,
signifying that the spin coherence is also affected anisotropically. Infact
it is seen from fig.2 that amplitude of oscillation is larger for the case
(i) and (ii), when 
the polarization vector  of Ferromagnets lies in 
{\it yz} or {\it xy}, compared to the case (iii). The absolute magnitude of
oscillation is always smaller than one implying even in ballistic transport
spin dephasing takes place due to the boundary scattering .

To further strengthen our results we performed numerical simulation on a
tight binding square lattice of lattice spacing {\it a} with 
$N_{x}$ sites along {\it x} axis and  $N_y$ sites along {\it y} axis. 
For tight binding Hamiltonian the Rashba spin-orbit coupling is
given by $\lambda_{so}=\alpha/2a=\tilde{\alpha}tk_{f}a/2$.
We fix $t$=1(hopping) and $k_{f}a$=1(ballistic case) 
for numerical simulation in tight binding model.
Once $t$ and $k_{f}a$ are fixed the other parameters for tight
binding model which would corresponds to the parameters of Fig. 1 are
given as, $N_{x}=2\pi\tilde{L}=50$,$N_{y}=2\pi\tilde{W}=50$ and
$\lambda_{so}=\tilde{\alpha}tk_{f}a/2=0.03$.With these set of parameters
we calculate spin resolved conductance for a given polarization
direction ($\theta, \phi$) of Ferromagnets, 
within Landauer-B\"uttiker formalism\cite{tribhu,tribhu1,but}. Using the spin resolved conductance
we define polarization as
\begin{equation}
P=\frac{G_{sc}-G_{sf}}{G_{sc}+G_{sf}}.
\label{eq9}
\end{equation}
where $G_{sc}$ and $G_{sf}$ are spin-conserved and spin flip conductance
respectively.
The quantity P in eq.(\ref{eq9}) corresponds to the quantity given in eq.(\ref{eq7}) and also
lies between +1 and -1.
This is plotted in Fig. 4 , we see that the agreement between Fig.3,
i.e, analytical calculation, and Fig. 4 is quite good. 
The slight quantitative mismatch is due to the fact that 
numerical simulation was done for
hard wall confining potential in {\it y} direction which leads to specular reflection, while in analytical calculation scattering from the boundary was
treated as diffusive. Therefore it is clear that the anisotropy
in spin transport is present in continuum model (effective mass Hamiltonian) 
as well as in tight binding model and is not an effect of reduced symmetry of
tight binding model, rather it is a consequence of breaking of rotational invariance due to spin orbit coupling. 

Now since conductance of FM/2DEG/FM, depends on the polarization of
electrons reaching the output terminal, hence it is expected  that conductance should also be anisotropic. 
This is clearly visible in Fig. 5 where we have plotted total conductance,
{\it i.e.}, $G=G_{sc}+G_{sf}$ corresponding to the Fig. 4,
as function of polarization angle.  It should be noted that the
conductance is symmetric with respect to angle $\theta$ or $\phi$
which is consistent with B\"uttiker symmetry relation for charge
transport\cite{but}. It is important to point out that in recent literature
\cite{grund}
an erroneous result was reported, where it was claimed that conductance of
a FM/2DEG interface changes on flipping the magnetization of FM  which is
incorrect.

The results presented above were in ballistic regime. To verify that
these results survives in diffusive case
we show polarization and conductance
in Fig. 5 and Fig. 6 respectively for diffusive case. We have taken
Anderson model for disorder with width 3$|t|$, corresponding to a mean free
path of $l=10a$. The other parameters are same as those for Fig. 3 and Fig. 4. 
It is clearly seen that
the anisotropy survives even in diffusive case.
This only strengthen our previous assertion that spin coherence is anisotropic.
Also it is instructive to compare Fig. 3 for ballistic transport and
Fig. 5 for diffusive transport. It is seen that the polarization which is transported is not affected much by the presence of disorder which is consistent with
the Rashba spin-orbit interaction which is independent of disorder strength.
However the magnitude of charge conductance is reduced drastically as
seen from Fig. 4 and Fig. 6, though the qualitative behavior as
function of angle remains unchanged. This clearly demonstrates that the
conductance anisotropy exist and is consistent with the B\"uttiker symmetry relation. One important thing to be noticed is the amplitude of oscillation for
ballistic case as well for diffusive case for both polarization and conductanceremains almost unchanged since the Rashba coupling was kept fixed for all the
figures. This clearly demonstrates that the anisotropy is a consequence of
spin-orbit interaction and is not affected by disorder.

I thank P. Bruno and G. Bouzerar for helpful discussion.

\newpage
\begin{center}
FIGURE CAPTINOS
\end{center}
\begin{description} 

\item{1}Fig.1 A 2DEG connected to two ideal Ferromagnetic leads.\label{Fig. 1}

\item{2}Fig2. Polarization as a function of angle calculated using eq.(\ref{eq7})
and eq.(\ref{eq8}).Where $\tilde{L}=\tilde{W}=50/2\pi$,$\tilde{\alpha}=0.06$.
\label{Fig. 2}

\item{3}Fig.3 Results of numerical simulation for polarization for 
ballistic system.
The numerical simulation were performed on a 50 $\times$ 50 lattice within tight binding model. The tight binding Rashba parameter is given by $\lambda_{so}
=\tilde{\alpha}tk_{f}a/2=0.03$, FM exchange
splitting is $\Delta/E_{f}=0.5$ and $k_{f}a$=1. These parameters were chosen in such a way that they correspond to the parameters of Fig. 1, as explained in text.\label{Fig. 3}

\item{4}Fig.4 The conductance as a function of angle. 
The parameters are same as
in Fig.3 \label{Fig. 4}

\item{5} Fig. 5 Polarization as a function of angle for diffusive case.
Here $k_{f}l$=10, where $l$ is mean free path.Configuration averaging was
performed over 15 different configuration. The other parameters are same
as in Fig. 3 \label{Fig. 5}

\item{6} Fig. 6 Conductance as function of angle corresponding to the Fig. 5.
Here $k_{f}l$=10, where $l$ is mean free path.Configuration averaging was
performed over 15 different configuration. The other parameters are same
as in Fig. 3 \label{Fig. 7}

\end{description}

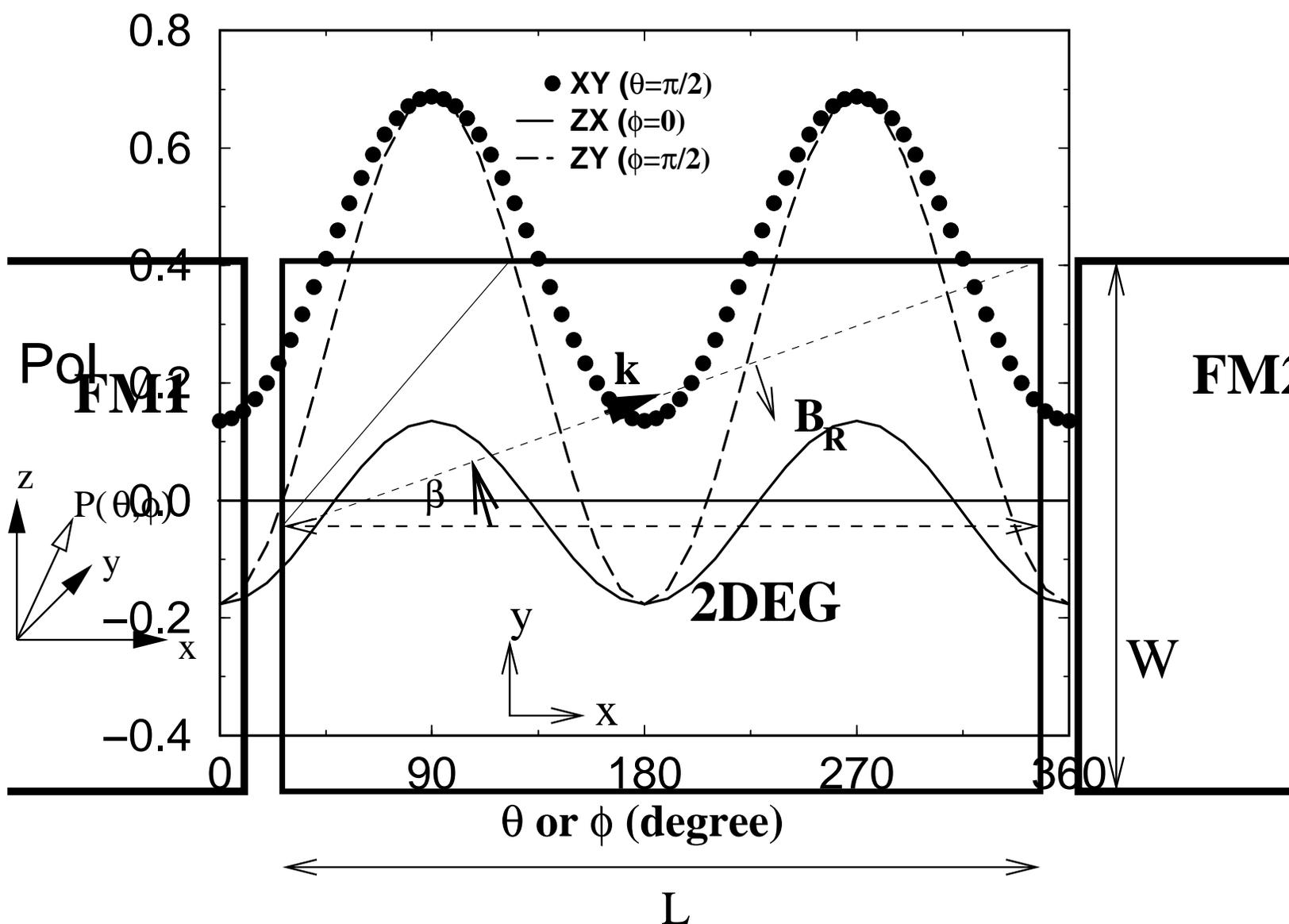
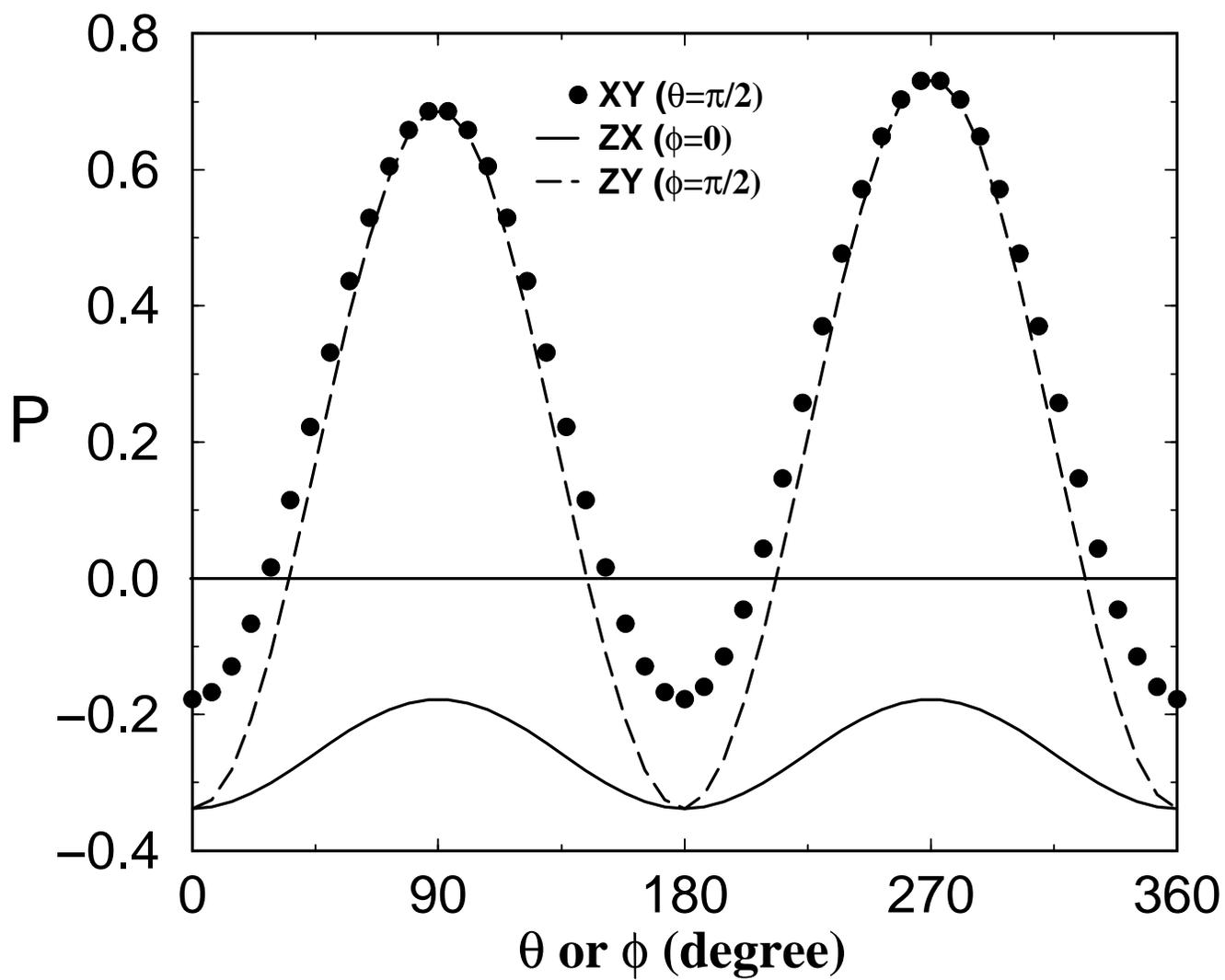
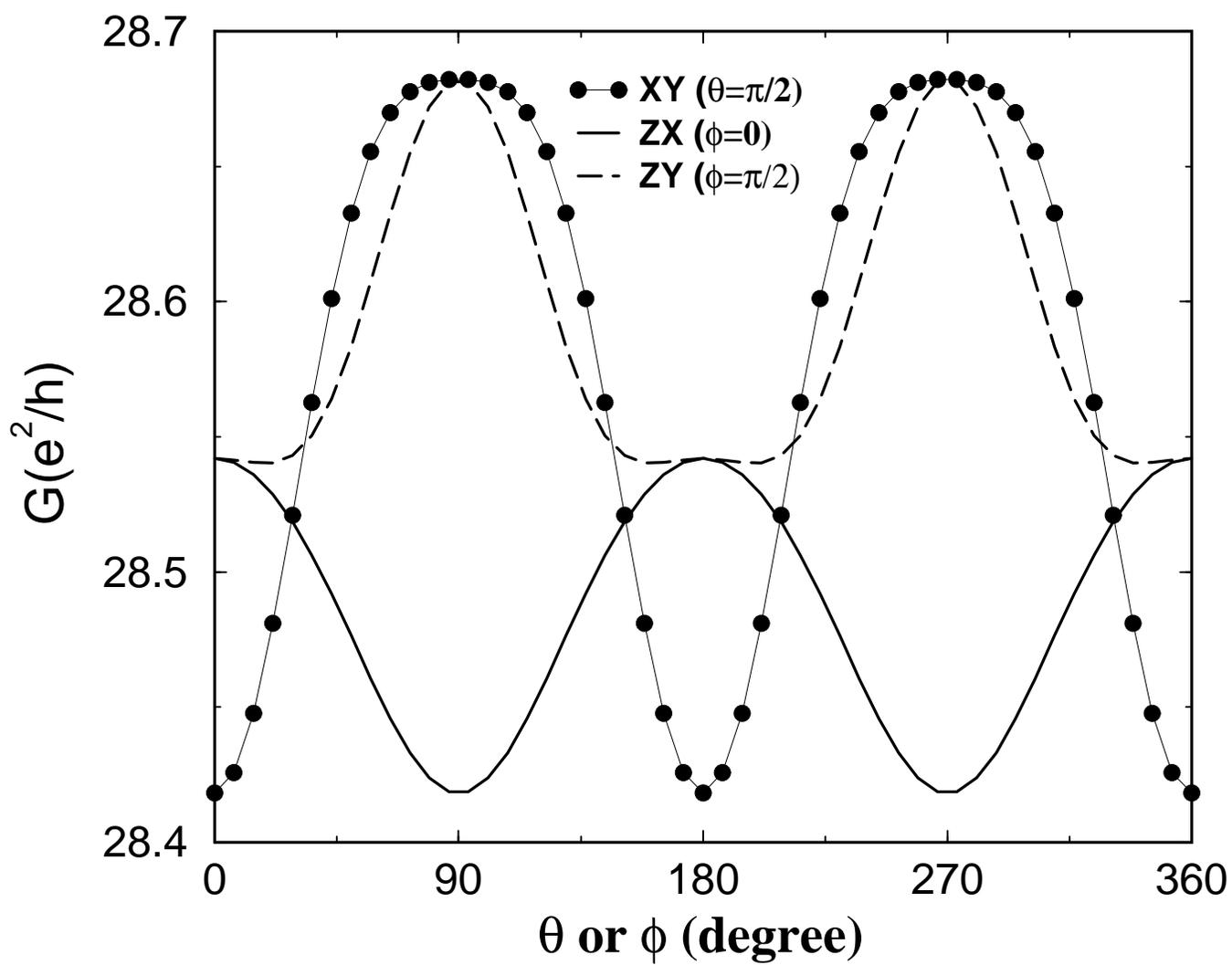
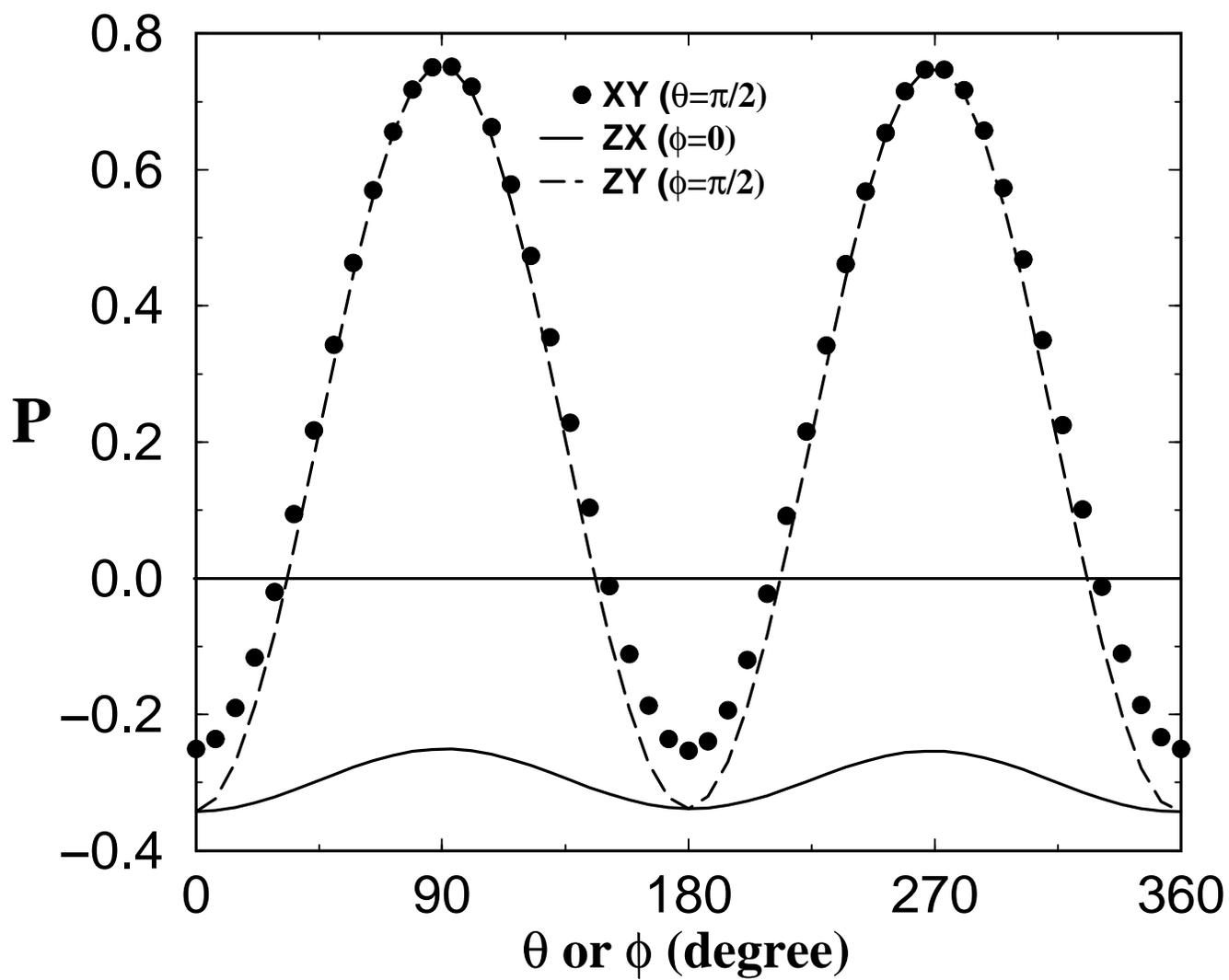
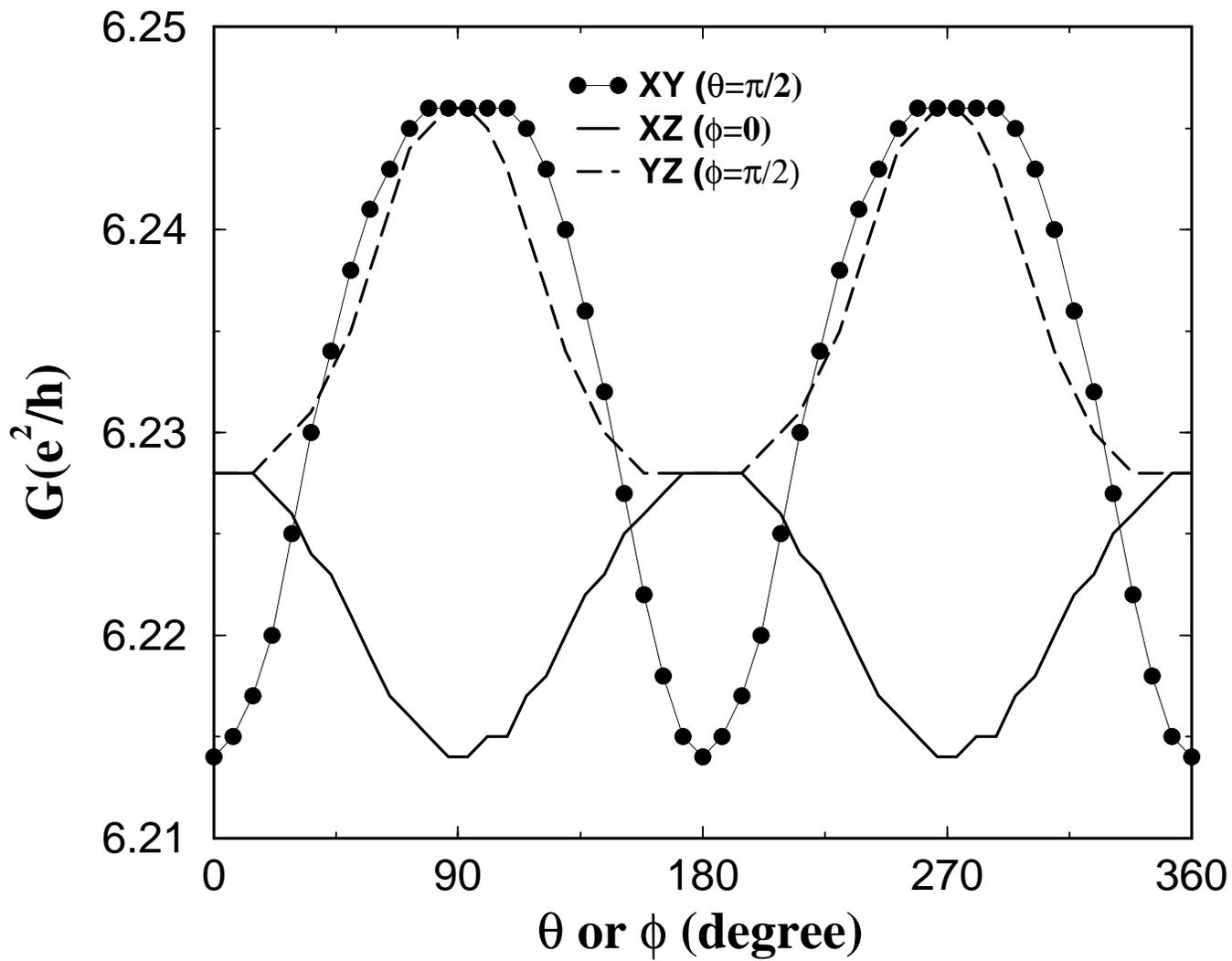
\end{document}